\documentclass{article}

\usepackage{graphicx} 
\usepackage{amsmath}   
\usepackage[compress]{cite}
\usepackage{amssymb} 
\usepackage{bm} 
\usepackage{dcolumn}
\usepackage{color}
\usepackage{mathrsfs}
\usepackage{amsfonts}
\usepackage{varioref}
 \usepackage{amsmath}
\usepackage{textcomp}

\RequirePackage[colorlinks,citecolor=blue,urlcolor=magenta,linkcolor=blue]{hyperref}
\allowdisplaybreaks

\labelformat{section}{Section #1} 
\labelformat{subsection}{Section #1} 
\labelformat{subsubsection}{Section #1}
\labelformat{subsubsubsection}{Section #1}
\labelformat{equation}{Eq.~(#1)} 
\labelformat{figure}{Fig.~#1} 
\labelformat{subfigure}{Fig.~\thefigure#1} 
\labelformat{table}{Table~#1} 
\labelformat{appendix}{Appendix #1}

\newcounter{mnotecount}[section]

\begin{document}
\title{GUP effects on Hawking temperature in a hot NUT Kerr Newman Kasuya Anti de Sitter black hole}
\author{ Heisnam Shanjit Singh \footnote{heisnam.singh@rgu.ac.in}$~^{1}$ and Khileswar Chandi \footnote{shanjitheisnam@gmail.com}$~^{1}$\\$^{1}$ \small{Department of Physics, Rajiv Gandhi University, Rono Hills, Doimukh Papumpare 791112, Arunachal Pradesh India}}

\date{\today}

\maketitle

\begin{abstract}
In this work, we use the generalised Klein-Gordon Equation in curved spacetime with an electromagnetic field to investigate the tunnelling phenomenon of scalar particles originating from the hot NUT Kerr Newman Kasuya Anti de Sitter (HNKNK AdS) black hole. Using the tunnelling formalism, we obtain a modified Hawking temperature different from previous works due to the quantum gravity effect for the charged Dirac particle at the HNKNK AdS black hole’s horizon. We find that the modified Hawking temperature is affected by the cosmological constant, angular momentum, magnetic mass, electric and magnetic charges. We demonstrate that a significant number of discontinuities exist in the heat capacities of the HNKNK AdS, indicating that the black hole system becomes unstable as the black hole size decreases.
\end{abstract}

\section{Introduction}\label{sec1}
Hawking \cite{Hawking1975} discovered the emission of thermal radiation, now called Hawking radiation, through a black hole's event horizon using quantum field theory in curved spacetime. Gibbons and Hawking \cite{Gibb} extended the research to cosmological event horizons and showed how quantum influences in spacetime lead to the emergence of thermal radiation.
The discovery of Hawking radiation has drawn significant attention as a potential pathway to quantum gravity as the Hawking radiation bridges statistical thermodynamics, quantum field theory and general relativity. To research the dynamical origins of the Hawking radiation, various methods like the tunnelling method \cite{Parikh, Kraus1}, the Hamilton-Jacobi method \cite{Angheben}, which extends Padmanabhan's complex path analysis \cite{Padhmanabhan} and the gravitational anomaly method \cite{Christensen}, emerge as a result of subsequent studies. For instance, Zhang and Zhao adopted the tunnelling method to Reissner-Nordström and Kerr-Newman black holes to research the emission of Hawking radiation \cite{zhang, zheng}).  An enhanced Hawking temperature from a Dyonic black hole was studied by applying the semi-classical approach, the Wentzel-Kramers-Brillouin (WKB) approximation and the Lagrangian equation in the presence of quantum gravity as seen in the generalised uncertainty principle (GUP)\cite{AliRi}. The modified Hawking temperature of the Schwarzschild black hole inside the bumblebee gravity model was addressed through the tunnelling probability by integrating the GUP-induced modification in the behaviour of the emitted particle into the Hamilton-Jacobi equation \cite{Sakallı}. Annamalai, D. and Pandey, A. \cite{Annamalai} investigated the Hawking radiation of a Schwarzschild-type black hole and the corrected Hawking temperature using the tunnelling method in Starobinsky-Bel-Robinson(SBR) gravity. Siahaan, Haryanto M., et al. \cite{Siahaan} used the tunnelling picture for the magnetised Taub NUT spacetime with the Manko–Ruiz parameters to compute Hawking radiation. The tunnelling phenomenon of a charged Dirac particle emerging from the thermal horizon of a hot NUT Kerr Newman Kasuya Anti de Sitter (HNKNK AdS) black hole was studied \ cite {shanjit}. The Hawking temperature and modified fermion tunnelling radiation of the stationary axisymmetric Kerr Taub NUT black hole in the curved spacetime were determined using the Lorentz dispersion relation correction \cite{Liu}. Employing gravitational anomalies, Murata and Soda \cite{Keiju} investigated Hawking radiation from rotating black holes and showed that the outer horizon radiates as a Schwarzschild black hole and the phenomenon results from spontaneous particle-antiparticle pair production near the event horizon.
In the literatures \cite{Ali, Juan,ak}, Hawking radiation was investigated through the tunnelling process for charged particles in complex spacetimes, such as Reissner Nordström Taub NUT and Kerr Newman black holes with electric and magnetic charges. Subsequently, a number of studies on Hawking radiation have appeared in the recent literature. For example, Media et al. \cite{Media:2024tyg} investigated the effects of Lorentz symmetry violation on the Hawking temperature and heat capacity of the KN AdS black hole surrounded by a perfect fluid. The radiation dynamics under GUP effects in a hot NUT Kerr Newman Kasuya Anti de Sitter black hole is yet to be explored. To investigate Hawking radiation utilising tunnelling methods, one determines the action for the emission process. Usually, the action becomes a complex quantity as the thermal emission across the event horizon is a classically forbidden process. It is well established that applying the WKB approximation, the tunnelling probability for the emission process is given by
\begin{eqnarray}
	\Gamma\sim exp[-\frac{2}{\hbar}Im I],
\end{eqnarray}
where I represents the classical action of the outgoing particle to leading order in $\hbar$ \cite{kerner1,Kerner2}.
We find that a GUP effect on the modified Hawking temperature of the Hot NUT Kerr Newman Kasuya Anti de Sitter black hole is affected by the mass, energy, angular momentum, charge and azimuthal angle of the emitted particle as well as the parameters of the black hole. Under the effects of GUP on scalar particle tunnelling, quantum gravity corrections slow down the rate of growth of the Hawking temperature for the black hole. Other black hole backgrounds are studied in \cite{wang}.\\
This paper investigates Hawking radiation and heat capacity for Dirac-charged particles emitted from the hot NUT Kerr Newman Kasuya Anti de Sitter (HNKNK-AdS) black hole via the tunneling process under GUP effect. This spacetime, characterised by parameters including mass, angular momentum per unit mass, the cosmological constant, and electric and magnetic charges, is particularly intriguing because it is theoretically predicted to contain magnetic monopoles. The investigation of quantum effects in the hot NUT Kerr Newman Kasuya Anti de Sitter (HNKNK-AdS) black hole is interesting because the fact that magnetic monopoles exist has been given on the grounds of the symmetry that they would introduce in the field equations of electromagnetism. We compute the tunnelling of a Dirac particle using the generalised Klein-Gordon equation to determine whether Hawking's temperature depends on the event horizon of the black hole.

\noindent The paper is organized as follows: In section 2, we review the HNKNK AdS black hole in curved spacetime. In section 3, we derive Hawking radiation using the Dirac equation under GUP effect and investigate global temperature for this spacetime. We also discuss the heat capacities of the HNKNK AdS black hole. In section 4, we discuss our findings. Throughout, we use natural units ($G = c = \hbar = 1$) unless specified otherwise. 

\section{Hot NUT Kerr Newman Kasuya Anti de Sitter spacetime}\label{sec2}

We consider an HNKNK AdS black hole characterized by the parameters: mass $ M$, specific angular momentum $ a = \frac{J}{M}$, NUT (magnetic mass) parameter $ n$, electric charge $ Q_e$, magnetic monopole parameter $ Q_m$ and the cosmological constant $ \Lambda$. According to an exact solution for a rotating dyon black hole, the metric for the four-dimensional HNKNK AdS black hole in Boyer-Lindquist coordinates is expressed as \cite{kasuya}:
\begin{equation}
	ds^2 = g_{tt} dt^2 + g_{rr} dr^2 + g_{\theta \theta} d\theta^2 + g_{\phi \phi} d\phi^2 + 2g_{t\phi} d\phi dt,
\end{equation}
where the metric components are given by:
\begin{equation}
	\begin{aligned}
		g_{tt} &= \frac{-\Delta + a^2 \Delta_\theta \sin^2\theta}{\Sigma},\\  \quad 
		g_{t\phi}& = g_{\phi t} = \frac{h \Delta - a \xi^2 \Delta_\theta \sin^2\theta}{\Xi \Sigma},\\ \quad 
		g_{\theta\theta}& = \frac{\Sigma}{\Delta_\theta}, \\
		g_{rr} &= \frac{\Sigma}{\Delta},\\ \quad 
		g_{\phi\phi} &= \frac{-h^2 \Delta+\xi^4 \Delta_\theta \sin^2\theta}{\Xi^2 \Sigma}.
	\end{aligned}
\end{equation}
The auxiliary functions are defined as:
\begin{equation}
	\begin{aligned}
		\Xi &= 1 - \frac{a^2}{y^2},\\ \quad 
		h & = a\sin^2\theta - 2n\cos\theta,\\ \quad 
		\Sigma &= r^2 + (n + a\cos\theta)^2,\\ \quad \xi^2 &= a^2 + r^2 + n^2, \\
		\Delta &= \xi^2 \left( \frac{5n^2 + r^2}{y^2} + 1 \right) - 2(Mr + n^2) + Q^2,\\ \quad 
		\Delta_\theta &= 1 - \frac{a^2 \cos^2\theta}{y^2}.
	\end{aligned} 
\end{equation}
Here, the negative cosmological constant is related to $ y$ by $ y^2 = -3/\Lambda$, and the total charge is $ Q^2 = Q_e^2 + Q_m^2$. The electromagnetic field tensor for this metric is:
\begin{eqnarray}
	F = \frac{Q}{\Sigma^2 \Xi} \left[ r^2 - (n + a\cos\theta)^2 \right] dr \wedge (dt - h d\phi)
	- \frac{2Qr}{\Sigma^2 \Xi} (n + a\cos\theta) \sin\theta \, d\theta \wedge (-a dt + \xi^2 d\phi).
\end{eqnarray}
The non-vanishing components of the electromagnetic potential are:
\begin{equation}
	A_\mu = -\frac{Qr}{\Xi \Sigma} \delta_{\mu t} + \frac{Qr h}{\Xi \Sigma} \delta_{\mu \phi}.
\end{equation}
The event horizons of the black hole are determined by the roots of $ \Delta = 0$. Near the event horizon $r=r_h$, $ \Delta$ can be expanded as:
\begin{equation}
	\Delta = (r - r_h) \frac{d\Delta}{dr}\Big|_{r=r_h} = (r - r_h) \Delta'\Big|_{r=r_h},
\end{equation}
where $ \Delta'$ denotes the derivative of $ \Delta$ with respect to $ r$ and $ r_h$ is a positive real root corresponding to the horizon. In the limiting cases: for $ y \to \infty$, $ Q_m = n = 0$, the metric reduces to the Kerr-Newman metric which exhibits axial singularities at $ \theta = 0$ and $ \theta = \pi$. For $ y \to \infty$, $ Q_e = Q_m = n = 0$, it reduces to the Kerr metric. For $ y \to \infty$, $ a = Q_m = n = 0$, it becomes the Reissner-Nordström metric. For $ y \to \infty$, $ a = Q_e = Q_m = n = 0$, it simplifies to the Schwarzschild metric.
\section{Generalised Klein-Gordon equation and Frame Dragging}
We consider, for a bound particle in a potential due to quantum fluctuations of background spacetime, the GUP-modified Dirac Hamiltonian obtained by Nozari et al. \cite{Nozari} as
\begin{eqnarray}
	\tilde{H} = (\boldsymbol{\alpha} \cdot \mathbf{p}) [1 - \beta(\nabla^2 + m^2)] + \beta_{\text{mat}} m, \label{nk}
\end{eqnarray}
where $\beta$ represents the strength of the Generalised Uncertainty Principle correction, $\alpha$ and $\beta_{mat}$ are $4\times 4$ Dirac matrices and $m$ is the mass of a particle. Now, we square up to the first order in $\beta$ and the second order in $p$, the Nozari-Karami Hamiltonian $\tilde{H}$ of equation (\ref{nk}) and obtain approximately as
\begin{eqnarray}
	\tilde{H}^2 &&\simeq p^2 + m^2 - q\boldsymbol{\Sigma} \cdot \mathbf{B} - 2\beta(p^2 + m^2)^2 + 2\beta q(p^2 + m^2)\boldsymbol{\Sigma} \cdot \mathbf{B},
\end{eqnarray}
where $\boldsymbol{\Sigma} \cdot \mathbf{B}$ with $\mathbf{B}=\mathbf{\nabla}\times \mathbf{A}$ is the spin-magnetic coupling which arises due to the interaction of the particle's spin and magnetic field. This coupling term vanishes in the case of the existence of the temporal component of the vector field only. For the particle with charge $q$ in a vector field: $A_{i}=(-\phi,\textbf{A})$ we use $p_{i}\rightarrow p_{i}-qA_{i}$. The generalised equation (\ref{nk}) in the presence of an electromagnetic field is written as
\begin{eqnarray}
	-g^{tt}\left(-i\hbar \partial_t - q A_t \right)\left( -i\hbar \partial_t - q A_t \right)\Psi&&=\left[ g^{\mu\nu}\left( -i\hbar \partial_\nu - q A_\nu \right)\left( -i\hbar \partial
	_\mu - q A_\mu \right) + m^2 \right]\times \nonumber \\
	&&\left[1-2\beta \left( g^{\mu\nu}\left( -i\hbar \partial_\mu-q A_\mu \right)\left( -i\hbar \partial_\nu - q A_\nu \right) + m^2 \right) \right]\Psi, \label{kg} 
\end{eqnarray}
where ${\mu,\nu=r,\theta,\phi}$ and $m$ is the mass of the test particle. An extremely rapidly rotating charged black hole drags inertial frames around it. Subsequently, photons and particles in the ergosphere are compelled to travel at an angular velocity around the black hole. The line element of the black hole has a singularity at the event horizon. The black hole has a frame-dragging effect in the coordinate system. By applying dragging coordinate transformation, $\frac{d\phi}{dt}=\frac{g_{t\phi}}{g_{\phi\phi}}=\omega$. The line element of the black hole in 3-dimensional space near the event horizon $r = r_h$ is given as
\begin{equation}
	ds^2 = g_{11}(r) dt^2 + g_{rr}(r) dr^2 + g_{\theta \theta}(r) d\theta^2,
\end{equation}
where 
\begin{equation}
	g_{11}(r)=g_{tt}(r)-\frac{g_{t\phi}(r)^2}{g_{\phi\phi}(r)}.
\end{equation}
The corresponding electromagnetic potential is given by transforming the vector potential 
$A_{\mu}$ into the new dragging frame where there is no longer a $d\phi$ term in the metric as
\begin{eqnarray}
	A_t= -\frac{Qr}{\Xi \Sigma}\left( \frac{g_{\phi\phi}+g_{t\phi}h}{g_{\phi\phi}} \right).
\end{eqnarray}
\section{Quantum tunneling from the hot NUT Kerr Newman Kasuya Anti de Sitter black hole}
We consider the tunneling of scalar particles from the black hole.The motion of scalar particles obey the generalized Klein–Gordon Equation (\ref{kg}). Using the inverse transformed metric, we get
\begin{eqnarray}
	&&g^{tt}\left(-i\hbar \partial_t - q A_t \right)^2\Psi=\left[ g^{rr}\left( -i\hbar \partial_r\right)\left( -i\hbar \partial
	_r\right)+g^{\theta\theta}\left( -i\hbar \partial_{\theta} \right)\left( -i\hbar \partial_{\theta}\right) + m^2 \right]\nonumber \\
	&&\times \left[ 1 - 2\beta \left( g^{rr}\left( -i\hbar \partial_r \right)\left( -i\hbar \partial_r\right)+g^{\theta\theta}\left( -i\hbar \partial_{\theta} \right)\left( -i\hbar \partial_{\theta}\right) + m^2 \right) \right] \Psi,
\end{eqnarray}
where $\Psi$ is given by
\begin{eqnarray}
	\Psi=exp[\frac{i}{\hbar}I(t,r,\theta)].
\end{eqnarray}
Using the WKB approximation, we obtain the Generalized Hamilton–Jacobi Equation as
\begin{eqnarray}
	&&-g^{tt}\left(-i\hbar \partial_t \frac{i}{\hbar}I - q A_t \right)^2=\left[ g^{rr}\left( -i\hbar \partial_r \frac{i}{\hbar}I\right)^2+g^{\theta\theta}\left( -i\hbar \partial_{\theta} \frac{i}{\hbar}I\right)^2 + m^2 \right]\times \nonumber \\
	&&\left[ 1 - 2\beta \left( g^{rr}\left( -i\hbar \partial_r \frac{i}{\hbar}I\right)^2+g^{\theta\theta}\left( -i\hbar \partial_{\theta} \frac{i}{\hbar}I\right)^2 + m^2 \right) \right], 
\end{eqnarray}
which reduces after simplification as
\begin{eqnarray}
	&&-g^{tt}\left(\partial_t I - q A_t \right)^2=\left[ g^{rr}\left(\partial_r I\right)^2+g^{\theta\theta}\left( \partial_{\theta} I\right)^2 + m^2 \right] \nonumber \\
	&&\times \left[ 1 - 2\beta \{ g^{rr}\left(\partial_r I\right)^2+g^{\theta\theta}\left( \partial_{\theta}I\right)^2 + m^2 \} \right].\nonumber
\end{eqnarray}
The action of the scalar particle is separable in the form of
\begin{eqnarray}
	I=-(\omega-\Omega L)t+R(r)+W(\theta)+\text{constant},
\end{eqnarray}
where $\omega$ is the energy of the particle, $L$ the angular momentum with respect to the angle $\phi$.
For a small value of $\beta$, we can write as follows:
\begin{eqnarray}
	&&g^{rr}\left(\partial_r I\right)^2+g^{\theta\theta}\left( \partial_{\theta} W\right)^2 + m^2=g^{tt}\left(\partial_t I - q A_t \right)^2+ 2\beta\left(g^{tt}\left(\partial_t I - q A_t \right)^2\right)^2.
\end{eqnarray}
Using Equation(\ref{kg}), we get
\begin{eqnarray}
	&&\partial_r R=\sqrt{g_{rr}g^{tt}}\sqrt{\left(\left(\omega-\Omega L+q A_t \right)^2-\frac{g^{\theta\theta}\left( \partial_{\theta} W\right)^2}{g^{tt}} - \frac{m^2}{g^{tt}}\right)}\nonumber \\
	&&\left(1+\frac{\beta\left(g^{tt}\left(\omega-\Omega L+q A_t+q A_t \right)^2\right)^2}{g^{tt}\left(\omega-\Omega L+q A_t \right)^2-g^{\theta\theta}\left( \partial_{\theta} W\right)^2 - m^2}\right).\nonumber
\end{eqnarray}
Integrating, we have
\begin{eqnarray}
	&&R_{\pm}=\pm\int\sqrt{g_{rr}g^{tt}}\sqrt{\left(\left(\omega-\Omega L+q A_t \right)^2-\frac{g^{\theta\theta}\left( \partial_{\theta} W\right)^2}{g^{tt}} - \frac{m^2}{g^{tt}}\right)}\nonumber \\
	&&\left(1+\frac{\beta\left(g^{tt}\left(\omega-\Omega L+q A_t+q A_t \right)^2\right)^2}{g^{tt}\left(\omega-\Omega L+q A_t \right)^2-g^{\theta\theta}\left( \partial_{\theta} W\right)^2 - m^2}\right)dr\nonumber,\\
	&&=\mp\frac{2\pi i\left(a^2+n^2+r_h^2\right)\left(\omega-j \Omega+q A_t\right)}{\Delta'(r_h)}\left[1+\frac{\beta}{2}\left(m^2+\frac{\left( \partial_{\theta} W\right)^2}{\Sigma(r_h) }\right)\right]+K, \nonumber \label{int}
\end{eqnarray}
where $K$ is a complex integration constant. Here, the $+$ sign represents the outgoing particle away from the black hole, whereas the $-$ sign represents the incoming particle towards the black hole. The integral of eqn. (\ref{int}) has only one pole at the event horizon $r_{h}$. Thus, the imaginary part of $\mathcal {R_\pm }$ is obtained as
\begin{eqnarray}
	\begin{aligned} 
		Im\mathcal{R_\pm}=\mp\frac{2\pi\left(a^2+n^2+r_h^2\right)\left(\omega-\omega_{0}\right)}{\Delta'(r_h)}
		\times \left[1+\frac{\beta}{2}\left(m^2+\frac{\left( \partial_{\theta} W\right)^2}{\Sigma(r_h) }\right)\right]-ImK,
	\end{aligned} \label{action}
\end{eqnarray}
where $\omega_{0}=j \Omega-q A_t$. The imaginary part of the action (\ref{action}) yields the imaginary part of the radial wave expression. Using the classical limit, the probability, $\Gamma$ of crossing the event horizon in each direction is proportional to 
\begin{eqnarray}
	&&\Gamma_{out}\propto\ e^{-Im\mathcal {I_\pm }}=e^{-(ImR_{+}+ImK)},\\
	&&\Gamma_{in}\propto e^{-Im\mathcal {I_\pm }}=e^{-(ImR_{-}+Im K)}. 
\end{eqnarray}
Since any incoming particle that crosses the event horizon has a 100\% chance of entering the black hole, the probability of a particle transitioning from outside to inside the horizon is equal to 1. This holds true only when $ \text{Im}R_{-} = -\text{Im}K$. Therefore, the probability of the particle tunnelling from inside to outside the horizon is given by:
\begin{eqnarray}
	\begin{aligned} \Gamma\propto\frac{\Gamma_{out}}{\Gamma_{in}}&=\frac{e^{-\Big(ImR_{+}+Im K\Big)}}{e^{-\Big(ImR_{-}+Im K\Big)}}=e^{-\Big(ImR_{+}-ImR_{-}\Big)},\\
		&=e^{-2ImR_{+}}.
	\end{aligned} \nonumber
\end{eqnarray}
It is assumed that the radiation is not purely thermal. It is known while employing the WKB approximation that the tunneling rate is related to the energy and the Hawking temperature of the relative particle as $\Gamma\propto e^{-\frac{\Delta\omega}{T}}$. If $\Delta\omega<0$ is the energy of the emitted particle, then the energy of the outgoing particle must be $-\Delta\omega$ due to the energy conservation. Thus, the probability of tunneling particle across the event horizon will read as $\Gamma\propto e^{\frac{\Delta\omega}{T}}=e^{\frac{\omega-\omega_{0}}{T}}$. Subsequently, the Hawking radiation denoted by $T_h$ is obtained as
\begin{eqnarray}
	T_h&&=\frac{\Delta'(r_h)}{4\pi\left(a^2+n^2+r_h^2\right)}\left[1+\frac{\beta}{2}\left(m^2+\frac{\left( \partial_{\theta} W\right)^2}{\Sigma(r_h) }\right)\right]^{-1}\nonumber\\
	&&\simeq\frac{\Delta'(r_h)}{4\pi\left(a^2+n^2+r_h^2\right)}\left[1-\frac{\beta}{2}\left(m^2+\frac{\left( \partial_{\theta} W\right)^2}{\Sigma(r_h) }\right)\right]\nonumber\\
	&&\simeq T_{0}\left[1-\frac{\beta}{2}\left(m^2+\frac{\left(\partial_{\theta} W\right)^2}{\Sigma(r_h) }\right)\right], \label{ht}
\end{eqnarray}
where 
\begin{eqnarray}
	&&T_{0}\simeq\frac{\Delta'(r_h)}{4\pi\left(a^2+n^2+r_h^2\right)}\nonumber \\
	&&\simeq\frac{r_h^2 \left(a^2+6 n^2+y^2\right)-a^2 \left(5 n^2+y^2\right)+3 r_h^4-5 n^4+n^2 y^2-Q^2 y^2}{4\pi\left(a^2+n^2+r_h^2\right)y^2 r_h} \nonumber
\end{eqnarray} 
is the usual Hawking temperature of the black hole without any quantum correction. The corrected Hawking temperature is lower than the original temperature as the corrected temperature depends on the mass of the emitted scalar particle, the azimuthal angle, and $\left(\partial_{\theta} I\right)^2$. The rate of increase of temperature during radiation is slowed down due to the quantum effect factor. In the absence of the quantum effect factor, the corrected Hawking temperature reduces to the original Hawking temperature. 

To get the specific heat, $C_{H}$ of the  HNKNK-AdS black hole, we obtain the black hole mass from $\Delta |_{r=r_h}=0$ as
\begin{eqnarray}
	M=\frac{r_h^2 \left(a^2+6 n^2+y^2\right)+a^2 \left(5 n^2+y^2\right)+r_h^4+5 n^4-n^2 y^2+Q^2 y^2}{2 y^2 r_h}.
\end{eqnarray}
Using the mass of the black hole, the heat capacity, $C_{H}$ of the black hole near the event horizon is obtained as\\
\begin{eqnarray}
	\begin{aligned}
		&&C_{H}=\Big(\frac{\partial M}{\partial r_{h}}\Big)\Big(\frac{\partial r_{h}}{\partial T_h}\Big)=-\frac{\pi A \left(a^2+r_h^2+n^2\right)^2 }{ B}\Big(1+\frac{\beta}{2}
		\Big(m^2+\frac{(\partial_{\theta} W)^2}{\Sigma}\Big)\Big),
	\end{aligned}
\end{eqnarray}
where 
\begin{eqnarray}
	A&=&-r_h^2\left(a^2+6n^2+y^2\right)+a^2 \left(5 n^2+y^2\right)-3 r_h^4+5 n^4-n^2y^2+Q^2y^2,\nonumber \\ 
	B&=&r_h^4 \left(8 a^2+3 n^2-y^2\right)+\left(a^2+n^2\right)\left(5a^2n^2+a^2y^2+5 n^4-n^2y^2+Q^2 y^2\right)\nonumber \\
	&&+r_h^2 \left(a^4+22a^2n^2+4a^2y^2+21n^4-2n^2y^2+3 Q^2 y^2\right)+3 r_h^6. \nonumber
\end{eqnarray}
\begin{figure}[htbp]
		\centering
		\includegraphics[width=.6\linewidth]{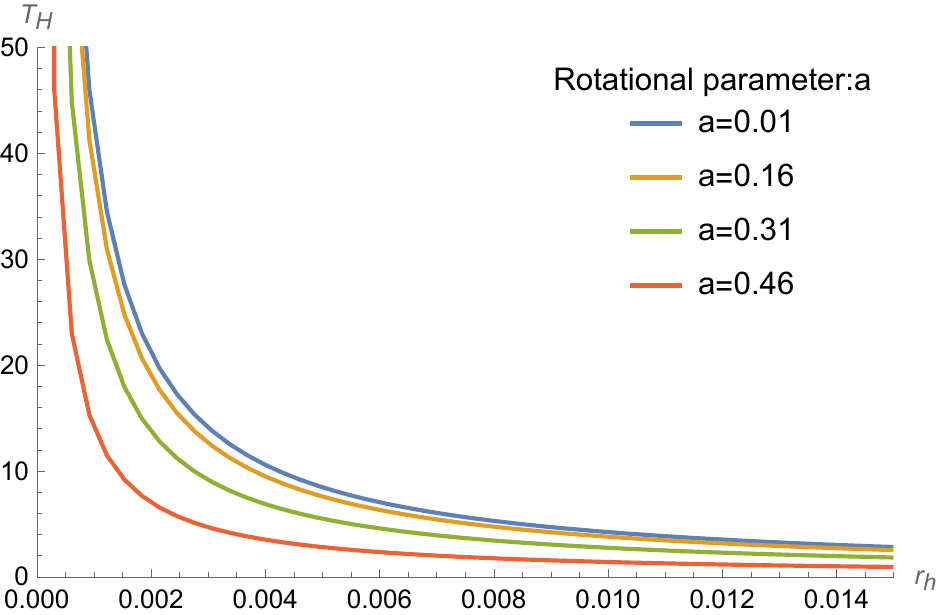}
		\caption{Variation of the Hawking temperature versus the various values of the event horizon of the black hole for different values of ``a".}
		\label{First}
	\end{figure}
	\hfill
	\begin{figure}
		\centering
		\includegraphics[width=.6\linewidth]{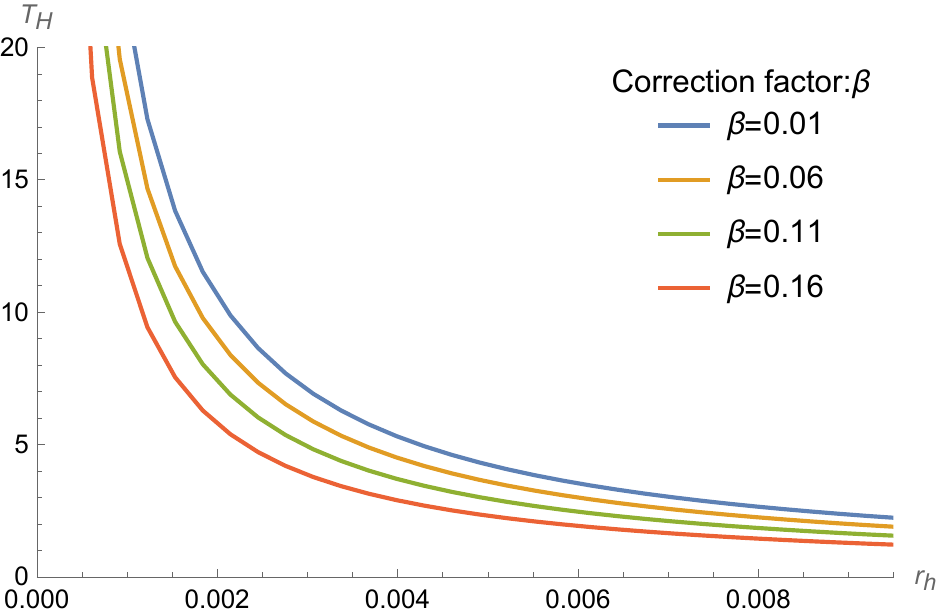}
		\caption{Variation of the Hawking temperature versus the various values of the event horizon of the black hole for different values of ``$\beta$".}
		\label{Second}
	\end{figure}
	\hfill
	\begin{figure}
		\centering
		\includegraphics[width=.6\linewidth]{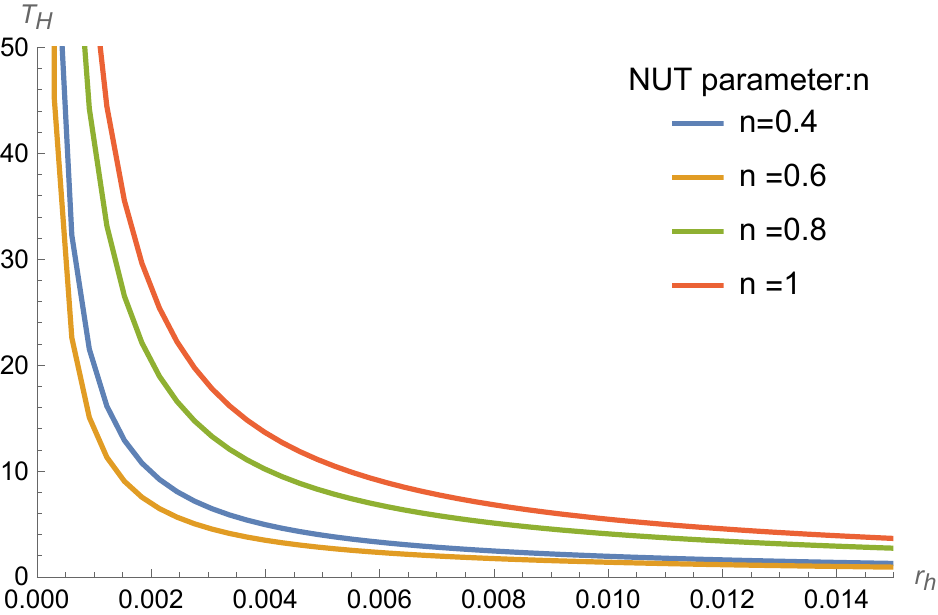}
		\caption{Variation of the Hawking temperature versus the various values of the event horizon of the black hole for different values of ``$n$".}
		\label{Third}
	\end{figure}
	\hfill
	\begin{figure}
		\centering
		\includegraphics[width=.6\linewidth]{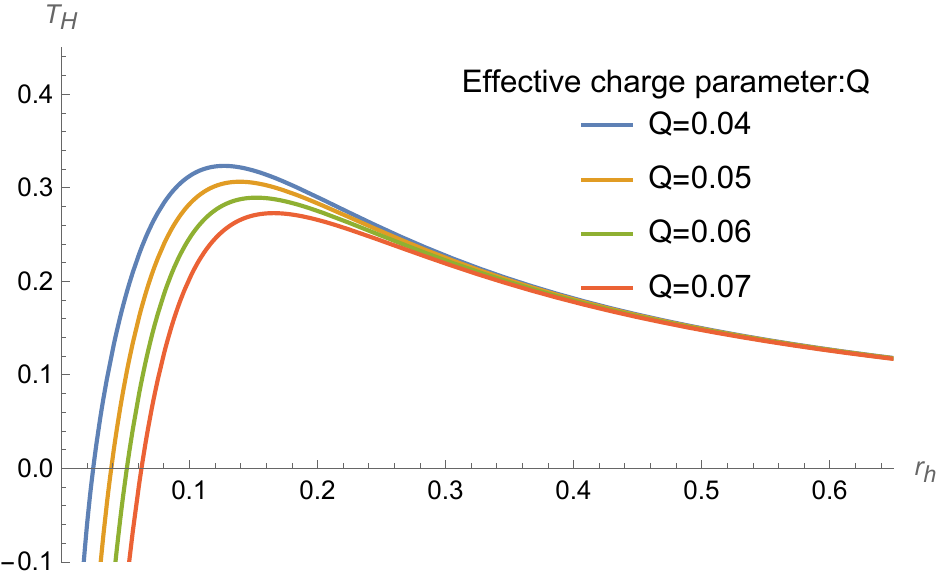}
		\caption{Variation of the Hawking temperature versus the various values of the event horizon of the black hole for different values of ``$Q$".}
		\label{Fourth}
	\end{figure}
Figure (\ref{First}) shows the Hawking temperature $T_H$ decreasing monotonically with increasing horizon radius $r_h$ for all values of the rotational parameter $a$. For a fixed $r_h$, larger rotation $a$ corresponds to a lower temperature, indicating that rotation suppresses $T_H$. The effect of rotation is strongest at small $r_h$, while all curves gradually converge at larger $r_h$, where the influence of a becomes weak. Figure (\ref{Second}) shows the Hawking temperature $T_H$ as a function of the horizon radius $r_h$ for different values of the correction factor $\beta$, where $T_H$ decreases monotonically with increasing $r_h$. For a fixed $r_h$, larger $\beta$ values lead to a lower Hawking temperature, indicating that the correction term suppresses $T_H$, especially at small horizon radii. Figure (\ref{Third}) shows that the Hawking temperature $T_H$ decreases monotonically with increasing horizon radius $r_h$ for all values of the NUT parameter n, diverging sharply as $r_{h}\rightarrow 0$. For a fixed $r_h$, larger values of $n$ correspond to higher temperatures, indicating that increasing the NUT parameter enhances the thermal profile of the black hole. In the figure(\ref{Fourth}), it is obvious that as $Q$ increases, the peak value of the Hawking temperature decreases and shifts slightly, indicating that a higher effective charge suppresses the maximum temperature while preserving a similar overall trend.

\begin{figure}[htbp]
		\centering
		\includegraphics[width=.6\linewidth]{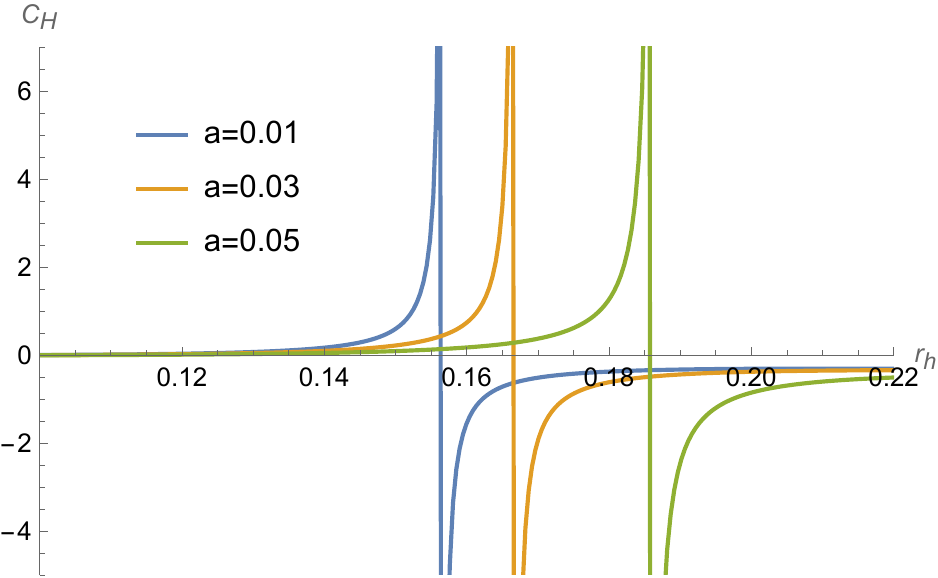}
		\caption{Variation of heat capacity versus the horizon radius for various values of ``a"}
		\label{heata}
	\end{figure}
	\hfill
	\begin{figure}
		\centering
		\includegraphics[width=.6\linewidth]{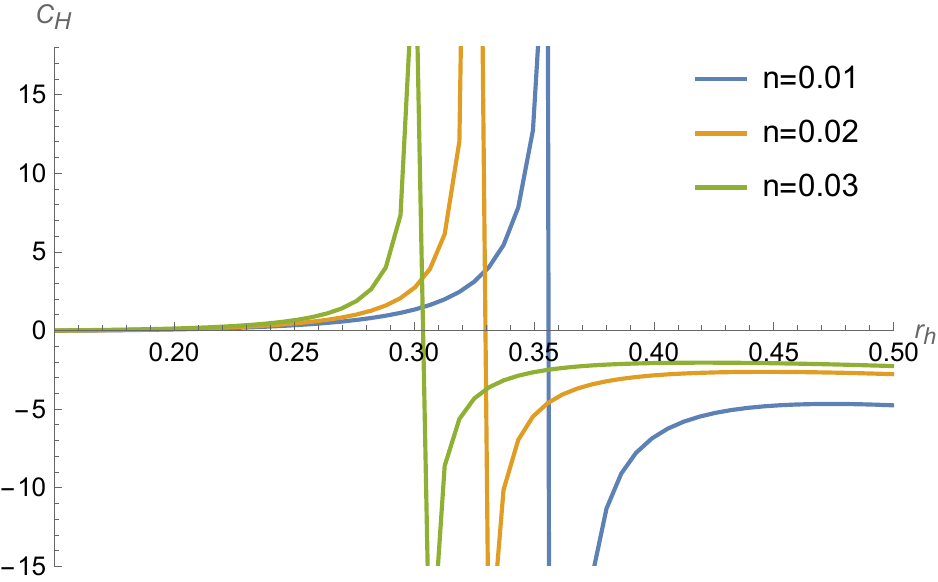}
		\caption{Variation of the Hawking temperature versus the various values of the event horizon of the black hole for different values of ``$n$".}
		\label{heatn}
	\end{figure}
	\hfill
	\begin{figure}
		\centering
		\includegraphics[width=.6\linewidth]{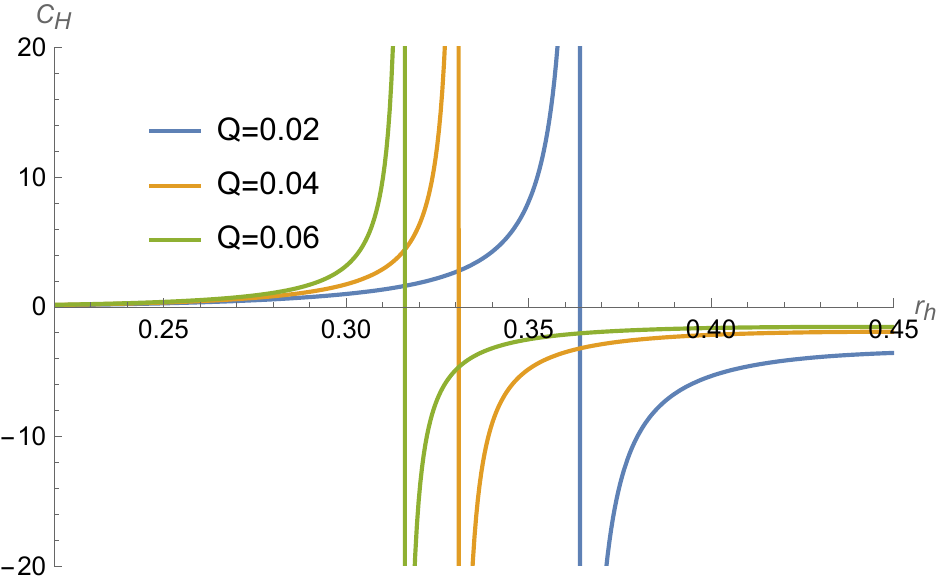}
		\caption{Variation of the Hawking temperature versus the various values of the event horizon of the black hole for different values of ``$Q$".}
		\label{heatQ}
	\end{figure}
	\hfill
	\begin{figure}
		\centering
		\includegraphics[width=.6\linewidth]{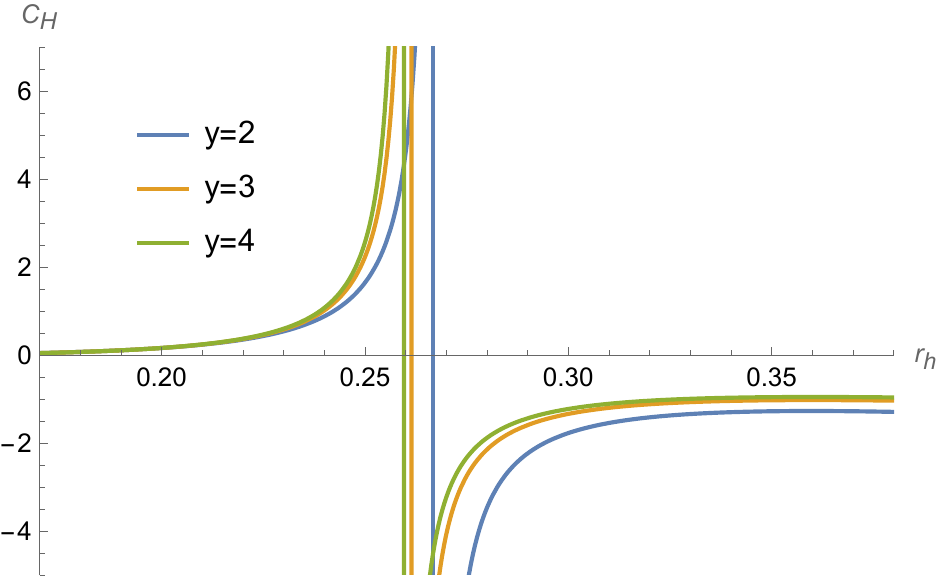}
		\caption{Variation of the Hawking temperature versus the various values of the event horizon of the black hole for different values of ``$y$".}
		\label{heaty}
	\end{figure}
	
Figure (\ref{heata}) shows that the black hole heat capacity $C_H$ diverges at a critical horizon radius $r_hr$, signalling a thermodynamic phase transition between stable $(C_H>0)$ and unstable $(C_H<0)$ black hole configurations. As the parameter $a$ increases, the divergence point shifts to larger $r_h$, indicating that the black hole’s thermodynamic stability and phase structure are strongly influenced by $a$. The figure (\ref{heatn}) shows how the quantity $C_H$ varies with radius $r$ for three different values of the parameter $n (0.01, 0.02, \text{and} 0.03)$. Each curve exhibits a sharp divergence (vertical asymptote), indicating a critical radius where $C_H$ changes sign, which shifts to smaller $r$ as $n$ increases. This behaviour suggests that increasing $n$ moves the critical point inward and alters the stability or phase behaviour of the system. The figure (\ref{heatQ}) shows that increasing $Q$ (from $0.02$ to $0.06$) shifts the divergence of $C_H$ to lower values of $r_h$, indicating that the critical (phase transition) radius decreases as charge increases. Additionally, the magnitude and sharpness of the divergence grow with $Q$, suggesting stronger thermodynamic instability at higher charge, while the figure (\ref{heaty}) shows the behaviour of $C_H$ as a function of the horizon radius $r_h$ for different values of the parameter $y$ $(y=2,3,4)$, where $C_H$ diverges at a critical $r_h$, indicating a phase transition. The change of sign of $C_H$ across this divergence suggests a transition between thermodynamically stable and unstable regimes, with the critical point shifting slightly as $y$ increases.

\section{Discussion and conclusion}\label{sec6}
In this work, we have investigated Hawking radiation from a hot NUT Kerr Newman Kasuya Anti de Sitter black hole using the tunnelling formalism in the presence of quantum gravity effects described by the Generalised Uncertainty Principle (GUP). By employing the generalised Klein–Gordon equation in curved spacetime with an electromagnetic field, we derived the GUP-corrected Hawking temperature for scalar particle emission across the event horizon. The corrected Hawking temperature is found to depend explicitly on the black hole parameters, including the angular momentum, NUT charge, electric and magnetic charges, and the cosmological constant, as well as on the mass and angular momentum of the emitted particle. The GUP correction introduces a negative contribution to the temperature, indicating that quantum gravity effects suppress Hawking radiation. Consequently, the rate of increase of the temperature during evaporation is slowed down, in agreement with earlier studies suggesting the possibility of long-lived black hole remnants. We also analyzed the thermodynamic stability of the HNKNK AdS black hole by evaluating the heat capacity near the event horizon. The heat capacity exhibits multiple divergences, signaling second-order phase transitions between stable and unstable configurations. These discontinuities become more pronounced for smaller horizon radii, indicating that small black holes are thermodynamically unstable, while larger black holes tend to evolve toward stable regimes. The locations of the critical points are strongly influenced by the rotation parameter, NUT charge, total charge, and the AdS curvature scale. In various limiting cases, our results reduce smoothly to those of Kerr–Newman, Kerr, Reissner–Nordström, and Schwarzschild black holes, confirming the consistency of the formalism. Overall, the present analysis demonstrates that the combined effects of rotation, gravitomagnetic mass, electromagnetic charges, and quantum gravity corrections play a crucial role in determining the thermal and stability properties of the HNKNK AdS black hole. These results provide further insight into Hawking radiation in complex spacetimes and may contribute to a deeper understanding of quantum gravity effects in black hole thermodynamics. The results suggest that the HNKNK AdS black hole provides a rich theoretical laboratory for probing quantum gravity effects in highly non-trivial spacetimes. Future investigations could extend this analysis to fermionic or vector particle tunnelling, explore higher-order GUP corrections, or examine the implications for black hole phase transitions in extended thermodynamic phase space. Such studies may offer deeper insight into the microscopic structure of spacetime and the end-point of black hole evaporation.

\section*{Acknowledgements}

The first author acknowledges IUCAA, Pune, for providing local hospitality and computational facilities to carry out this research work. The second author acknowledges the authority of Rajiv Gandhi University to provide the necessary facilities and support to carry out this work. Both authors gratefully acknowledge the support of the Anusandhan National Research Foundation (ANRF), Government of India, for funding and facilitating this project work. The assistance provided through the ANRF grant no.: ANRF/IRG/2024/000110/PS played a crucial role in conducting this research.\\

\end{document}